\documentclass[10pt,twocolumn,letterpaper]{article}

\usepackage{cvpr}              % To produce the CAMERA-READY version
\definecolor{cvprblue}{rgb}{0.21,0.49,0.74}
\usepackage[pagebackref,breaklinks,colorlinks,allcolors=cvprblue]{hyperref}
\usepackage{xcolor}
\usepackage{dsfont}
\usepackage{amsmath,amsfonts}
\usepackage{algorithmic}
\usepackage{algorithm}
\usepackage{lscape}
\usepackage{array}
\usepackage{textcomp}
\usepackage{stfloats}
\usepackage{url}
\usepackage{float}
\usepackage{booktabs}
\usepackage{verbatim}
\usepackage{graphicx}
\usepackage{cite}
\usepackage{colortbl,hhline}
\usepackage{subcaption}
\usepackage{multirow}
\usepackage{multicol}
\usepackage{pifont}
\usepackage[flushleft]{threeparttable}
\newcommand{\cmark}{\ding{51}}
\newcommand{\xmark}{\ding{55}}

\DeclareRobustCommand{\IEEEauthorrefmark}[1]{\smash{\textsuperscript{\footnotesize #1}}}

\def\paperID{*****} % *** Enter the Paper ID here
\def\confName{CVPR}
\def\confYear{2026}

\title{StrokeNeXt: A Siamese-encoder Approach for Brain Stroke Classification in Computed Tomography Imagery}

\author{Leo Thomas Ramos\IEEEauthorrefmark{1,2}\qquad 
    Angel D. Sappa\IEEEauthorrefmark{1,2,3} \\
    \IEEEauthorrefmark{1}Computer Vision Center \quad \IEEEauthorrefmark{2}Universitat Autònoma de Barcelona \quad \IEEEauthorrefmark{3}ESPOL Polytechnic University\\
}

\begin{document}
\maketitle
\begin{abstract}
We present StrokeNeXt, a model for stroke classification in 2D Computed Tomography (CT) images. StrokeNeXt employs a dual-branch design with two ConvNeXt encoders, whose features are fused through a lightweight convolutional decoder based on stacked 1D operations, including a bottleneck projection and transformation layers, and a compact classification head. The model is evaluated on a curated dataset of 6,774 CT images, addressing both stroke detection and subtype classification between ischemic and hemorrhage cases. StrokeNeXt consistently outperforms convolutional and Transformer-based baselines, reaching accuracies and F1-scores of up to 0.988. Paired statistical tests confirm that the performance gains are statistically significant, while class-wise sensitivity and specificity demonstrate robust behavior across diagnostic categories. Calibration analysis shows reduced prediction error compared to competing methods, and confusion matrix results indicate low misclassification rates. In addition, the model exhibits low inference time and fast convergence. Code is available at: \url{www.hidden_for_review.com}
\end{abstract}
%\url{https://github.com/Leo-Thomas/StrokeNeXt}    
\section{Introduction}
\label{sec:intro}

Brain stroke is a life-threatening medical condition and a leading cause of adult mortality and long-term disability worldwide, affecting millions of individuals each year \citep{ZHU2022147}. It occurs when cerebral blood flow is disrupted due to vascular blockage or rupture \citep{HOSSAIN2025109711,10112284}, depriving brain tissue of oxygen and nutrients and resulting in cellular injury or death \citep{HOSSAIN2025109711}. Owing to its abrupt onset and severe clinical consequences, stroke remains a major challenge in emergency medicine and neurology \citep{GAUTAM2021102178}.

Clinically, strokes are broadly categorized into ischemic and hemorrhagic types, depending on whether they are caused by vascular blockage or rupture \citep{HOSSAIN2025109711}. Typically, initial assessment relies on neuroimaging techniques such as Magnetic Resonance Imaging (MRI) and Computed Tomography (CT) \citep{cabral_powers_2022,ABBASI2023100145}, which enable visualization of affected brain regions and discrimination between stroke subtypes, thereby supporting timely therapeutic decisions \citep{ABBASI2023100145}.

However, image interpretation remains complex and highly time-sensitive \citep{DESHPANDE2021102573}. Manual analysis demands expert radiologists and often involves labor-intensive procedures to identify subtle lesions or distinguish pathological patterns \citep{ABBASI2023100145}, making it susceptible to human error and inter-observer variability \citep{AGGARWAL2022105350}. Also, high operational costs and the need for specialized personnel can restrict timely diagnosis in resource-limited environments \citep{GAUTAM2021102178,CAI2022102522}. These challenges have driven increasing interest in computational approaches, particularly Artificial Intelligence (AI), to support and enhance clinical decision-making in stroke assessment.

AI methods, particularly those based on computer vision, enable automated analysis of medical images by learning discriminative visual patterns from large annotated datasets \citep{ELHARROUSS2024100645,10623211}. In stroke diagnosis, such models can identify subtle abnormalities in MRI or CT scans with minimal human intervention, supporting more consistent and efficient assessment. Consequently, AI-based solutions have gained increasing adoption in clinical research and practice \citep{esteva_chou21}.

Given the critical nature of stroke assessment, effective methods must combine high diagnostic accuracy with computational efficiency \citep{YALCIN2022105941}. Misclassification of stroke subtypes can lead to inappropriate treatment decisions, increasing the risk of irreversible brain damage or death \citep{CAI2022102522}, while excessive computational cost may limit deployment in time-critical clinical settings \citep{abdou_2022}. Accordingly, balancing reliable predictions with fast inference is essential.

% Building upon the aforementioned challenges, this work presents StrokeNeXt (Fig. \ref{fig:overview}), a Deep Learning (DL) approach for brain stroke classification based on a dual-branch feature extraction architecture. Both branches process the same input image independently, allowing the model to capture complementary representations and reduce the risk of information loss associated with single-path pipelines. The extracted features are then processed by a lightweight yet effective decoding module designed to enhance feature fusion and produce the final classification. The proposed method is evaluated on a real-world dataset of CT scans, demonstrating competitive performance. 

Based on the aforementioned, we present StrokeNeXt (Fig. \ref{fig:overview}), a Deep Learning (DL) approach for brain stroke classification based on a dual-branch feature extraction architecture. The two branches process the same input image independently, enabling the capture of complementary representations and mitigating information loss typically associated with single-path pipelines. The resulting features are fused through a lightweight decoding module designed to enhance feature integration and generate the final classification. StrokeNeXt is evaluated on a real-world CT dataset, where it demonstrates competitive performance. The main contributions of this work are summarized as follows:

\begin{figure}[t]
    \centering
    \includegraphics[width=1\linewidth]{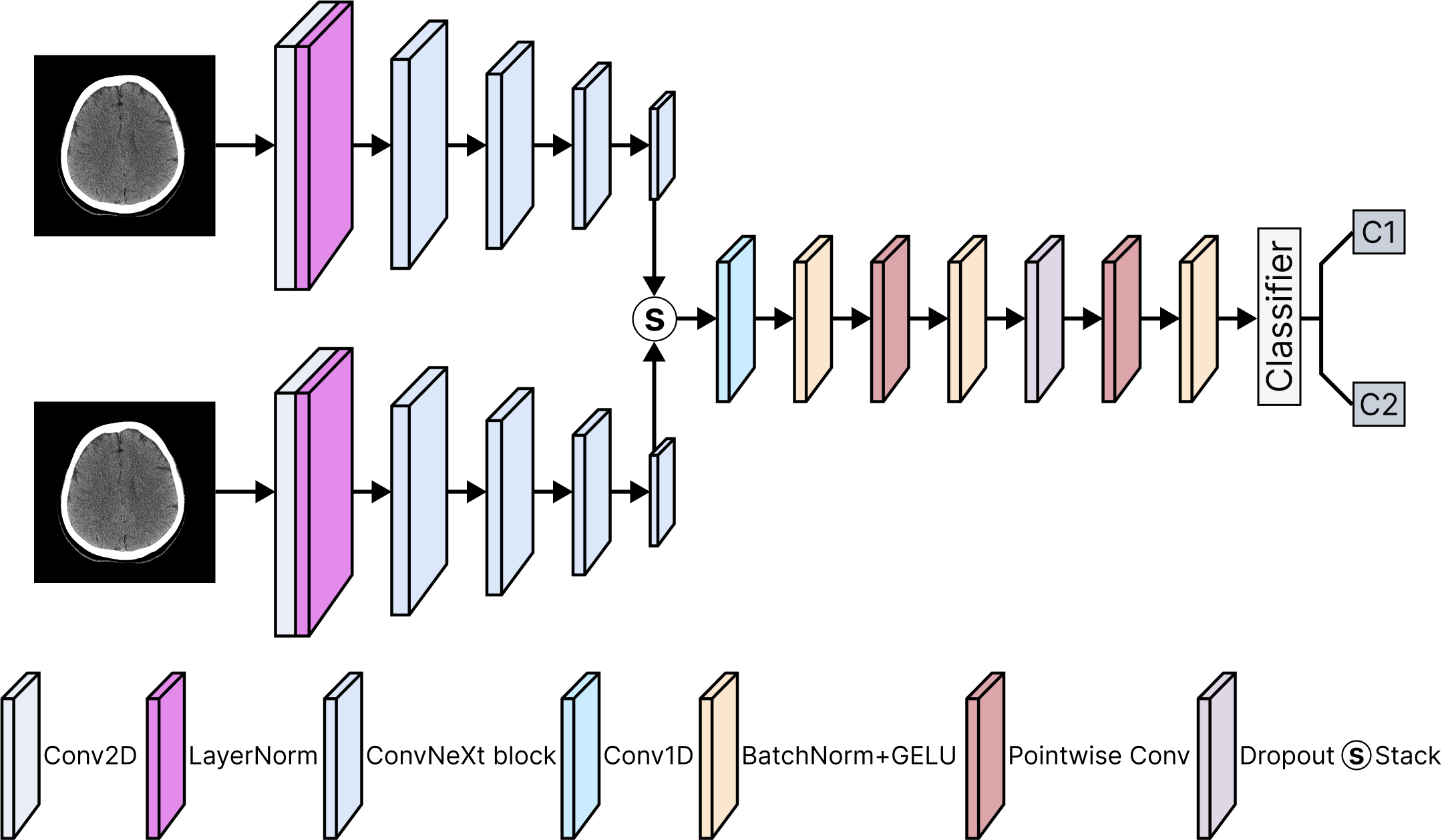}
    \caption{Overview of the proposed StrokeNeXt architecture. The model processes the input CT image through two parallel ConvNeXt encoders. Their outputs are fused via a custom convolutional decoder composed of a stack operation, a 1D convolutional fusion path with transformation layers, and a classification head.}
    \label{fig:overview}
\end{figure}

\begin{itemize}
    \item We present StrokeNeXt, an approach for brain stroke classification that integrates dual-branch feature extraction and a lightweight decoder.
    \item We demonstrate that StrokeNeXt achieves high accuracy in distinguishing between stroke and normal cases using real-world CT scan data.
    \item We show that the model also performs effectively in differentiating among various stroke subtypes.
    \item We show that StrokeNeXt performs notably when compared to other state-of-the-art models from the literature.
    \item We establish a baseline architecture and performance benchmark that can serve as reference for future research.
\end{itemize}

% The remainder of this article is organized as follows. Section \ref{sec:related} reviews related work on brain stroke classification using intelligent methods to contextualise our proposal. Section \ref{sec:II} introduces the methodology, describing the architectural design, dataset, and experimental protocol. Section \ref{sec:III} reports and discusses the results across all experiments performed. Section \ref{sec:IV} addresses the main limitations of the study. Lastly, Section \ref{sec:V} provides the conclusions and highlights possible avenues for future research.
\section{Related work}\label{sec:related}

Stroke identification has been widely explored using diverse paradigms. Early approaches relied on traditional classifiers such as Random Forest \citep{9179307} and ElasticNet \citep{DEV2022100032}, which achieved moderate success in stroke detection but were limited by their dependence on handcrafted features and lack of spatial context. Subsequent efforts shifted toward DL, where CNN-based models such as MobileNetV2 \citep{YALCIN2022105941} improved automation and spatial feature extraction but still exhibited limited representational capacity and robustness.

Other architectures have been proposed to enhance feature representation. Models such as P-CNN \citep{GAUTAM2021102178}, D-UNet \citep{YALCIN2022105941}, and OzNet \citep{bioengineering9120783} improve lesion localization through multi-scale encoding or task-specific optimizations. In parallel, 3D approaches, including 3D-CNN \citep{NEETHI2022103720} and CAD systems \citep{TURSYNOVA20231431}, exploit volumetric information to capture richer spatial context. However, these methods typically rely on computationally expensive 3D pipelines, increasing complexity and overfitting risk, limiting their suitability for real-time clinical deployment.

Recent work has explored hybrid architectures to model long-range dependencies. For example, StrokeViT \citep{RAJ2023105772} combines convolutional networks with Vision Transformers to enhance feature representation and prediction accuracy. However, Transformer-based models generally demand large datasets, high computational cost, and long training times, limiting their practicality in resource- and time-constrained clinical settings.

Overall, existing methods exhibit a trade-off between accuracy, efficiency, and reliability. While some achieve strong classification performance, they often incur high computational cost, whereas others lack proper calibration for time-sensitive clinical deployment. Addressing these limitations, StrokeNeXt proposes a dual-branch convolutional framework that strengthens feature diversity through parallel processing while maintaining efficiency. Its design focuses on achieving robust diagnostic performance without relying on large data volumes or computationally expensive inference pipelines, positioning it as a balanced alternative within the current landscape of stroke analysis methods.

\section{Methodology}\label{sec:II}

\subsection{Dataset and preprocessing}

\begin{figure}[b]
    \centering
    \includegraphics[width=0.7\linewidth]{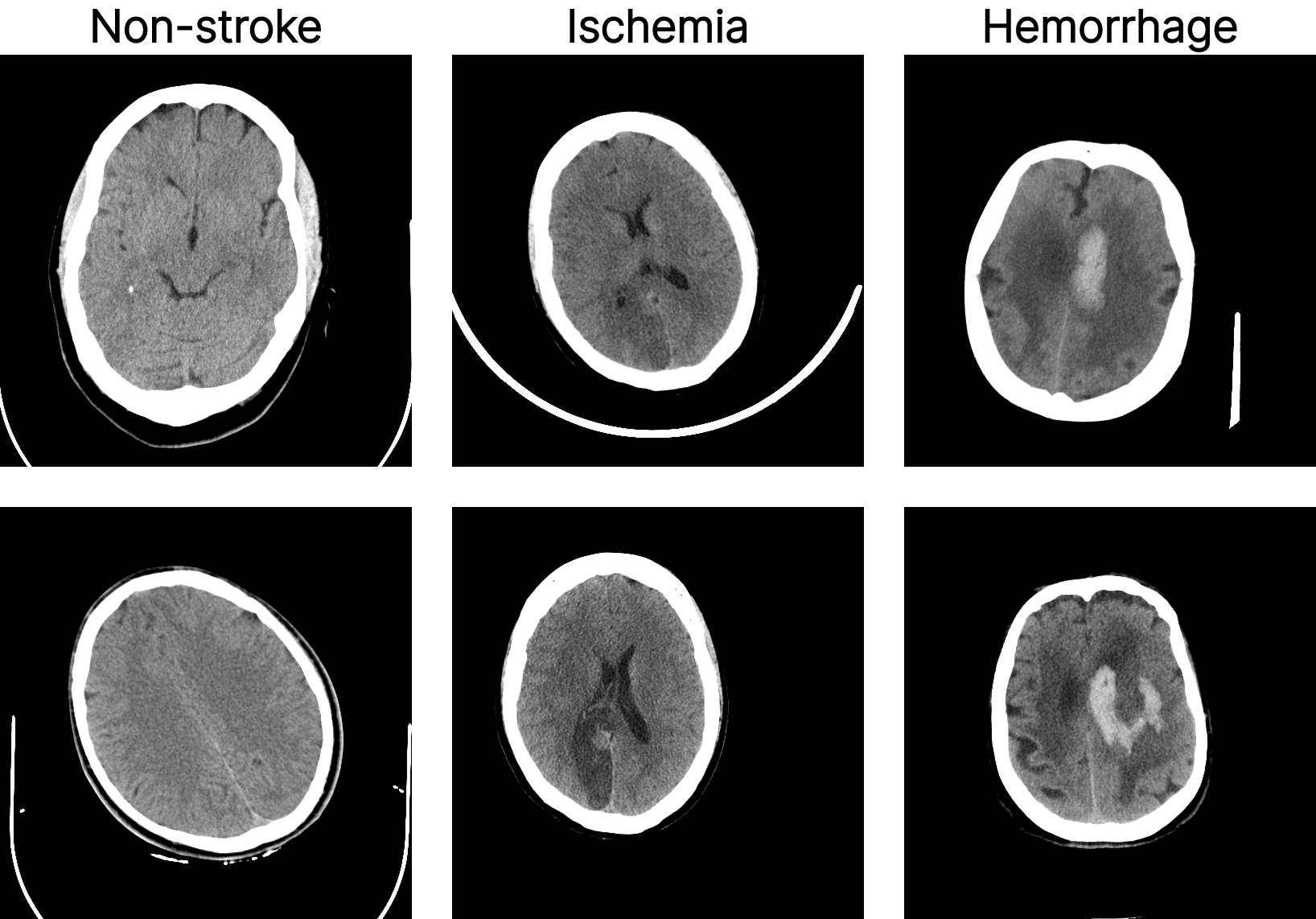}
    \caption{Sample images from the used dataset in this work.}
    \label{fig:dataset}
\end{figure}

We use the Stroke Dataset released as part of the Artificial Intelligence in Healthcare Competition (TEKNOFEST 2021) \citep{koc_akc2022}. It consists of 6,774 brain CT cross-sectional images in PNG format, annotated and curated by a team of seven radiologists. The dataset is divided into three classes: 4,551 non-stroke, 1,093 hemorrhagic stroke, and 1,130 ischemic stroke cases. The dataset is publicly available, and representative samples are shown in Fig. \ref{fig:dataset}.

To structure the experiments, the dataset was reorganized into two settings. The first addresses stroke detection by grouping ischemic and hemorrhagic cases into a single positive class and using non-stroke samples as negatives. The second focuses on stroke subtype classification, considering only ischemic and hemorrhagic cases and excluding non-stroke images. In both settings, the data were randomly split into training, validation, and test sets following an 80-10-10 ratio. Data augmentation was used, including random horizontal flips, rotations up to 10°, and brightness, contrast, and saturation jittering. Images were resized to 224$\times$224 and normalized using standard ImageNet statistics.

\subsection{Model design}

\subsubsection{Siamese encoder}

% Our model incorporates a dual-branch feature extraction module, where two identical encoders process the same input image in parallel. The underlying hypothesis is that two independent branches can explore different abstraction paths even from the same input, because the parameters are not shared and optimization can lead them to represent different aspects of the image. In other words, we argue that parallel paths increase representational diversity because each branch follows its own stochastic trajectory during training, which naturally pushes the filters to specialize on different visual cues. This helps improve the features passed to the decoder, since each encoder may provide complementary features that simple single-branch extractors might not capture.

StrokeNeXt adopts a dual-branch feature extraction module in which two identical encoders process the same input image in parallel. The core hypothesis is that independent branches can explore different abstraction paths despite sharing the same input, since their parameters are not shared and optimization drives them toward distinct representations. As each branch follows its own stochastic training trajectory, the encoders naturally specialize in different visual cues, increasing representational diversity. This results in complementary features being passed to the decoder, improving feature quality compared to single-branch extractors.

Each encoder branch is based on ConvNeXt \citep{convnext}, a CNN architecture inspired by Transformer design principles. ConvNeXt is built on the hypothesis that fully convolutional models can match or surpass Transformer performance without incurring the high computational and data demands typically associated with it. ConvNeXt has been validated across multiple benchmarks, motivating its use as the backbone of the proposed dual-branch encoder, where it provides strong representational capacity while preserving computational efficiency for stroke classification. %typically associated with attention-based architectures.

% ConvNeXt builds upon the ResNet family with key modifications such as larger kernel sizes (7$\times$7), GELU activations, layer normalization, and a stage-wise structure. This makes ConvNeXt effective for hierarchical representation learning with controlled computational cost. Structurally, ConvNeXt consists of four sequential stages of convolutional blocks followed by a pooling layer and a classification head, as shown in Fig. \ref{fig:encoder}. In our implementation, we remove the final pooling and classifier layers, retaining only the feature extraction stages to serve as the encoder component of each branch. The architecture is available in four variants: tiny, small, base, and large. The primary structural difference among these variants is their depth and width, specifically the number of channels and the number of ConvNeXt blocks within each stage of the network. Details of these models can be seen in Table \ref{tab:conv_variants}. 

\begin{figure}[!ht]
    \centering
    \includegraphics[width=\linewidth]{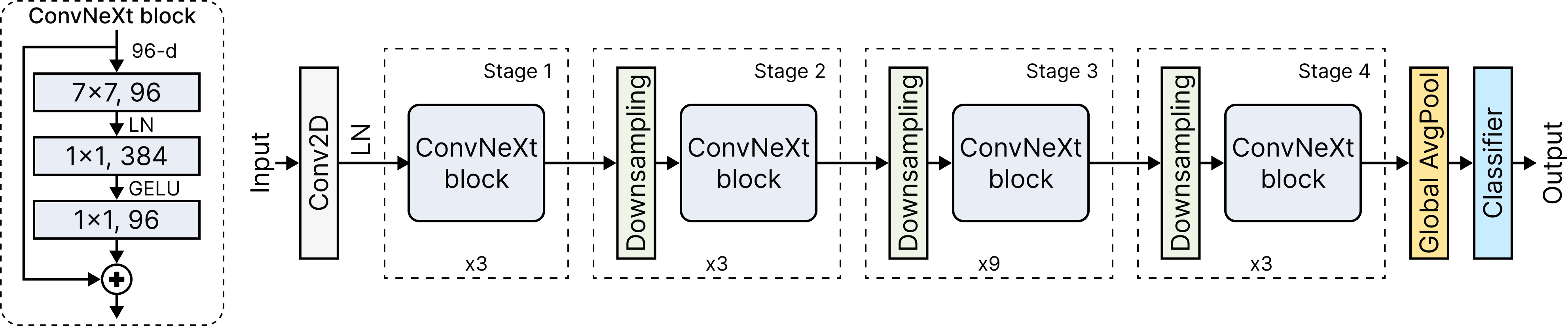}
    \caption{ConvNeXt architecture (tiny version). Source: \citep{10971457}}
    \label{fig:encoder}
\end{figure}

ConvNeXt builds upon a ResNet50 through key modifications, including larger kernel sizes (7$\times$7), GELU activations, layer normalization, and a stage-wise design, enabling effective hierarchical representation learning with controlled computational cost. Architecturally, it comprises four sequential stages of convolutional blocks followed by pooling and a classification head, as illustrated in Fig. \ref{fig:encoder}. In our implementation, the final pooling and classifier layers are removed, and only the feature extraction stages are retained to serve as the encoders. ConvNeXt is available in tiny, small, base, and large variants, which differ primarily in network depth and width, namely the number of channels and blocks per stage, as summarized in Table \ref{tab:conv_variants}.

\begin{table}[ht!]
\centering
\caption{Details of ConvNeXt variants. Taken from: \citep{ramos2026multiencoder}}\label{tab:conv_variants}
\resizebox{\linewidth}{!}{%
\begin{tabular}{p{2.5cm}p{3cm}p{2.5cm}p{1.65cm}p{1.5cm}}
\toprule
\textbf{Model} & \textbf{Channels {\scriptsize(per stage)}} & \textbf{Depths {\scriptsize(per stage)}} &\textbf{Params {\scriptsize (M)}} & \textbf{FLOPs {\scriptsize  (G)}}\\ \midrule
ConvNeXt-tiny  & [96, 192, 384, 768] & [3, 3, 9, 3] & 28 & 4.5\\
ConvNeXt-small  & [96, 192, 384, 768] & [3, 3, 27, 3] & 50 & 8.7\\
ConvNeXt-base  & [128, 256, 512, 1024] & [3, 3, 27, 3] & 89 & 15.4\\
ConvNeXt-large  & [192, 384, 768, 1536] & [3, 3, 27, 3] & 198 & 34.4\\
\bottomrule
\end{tabular}}
\end{table}

\subsubsection{Fusion decoder}

\begin{figure}[!ht]
    \centering
    \includegraphics[width=0.65\linewidth]{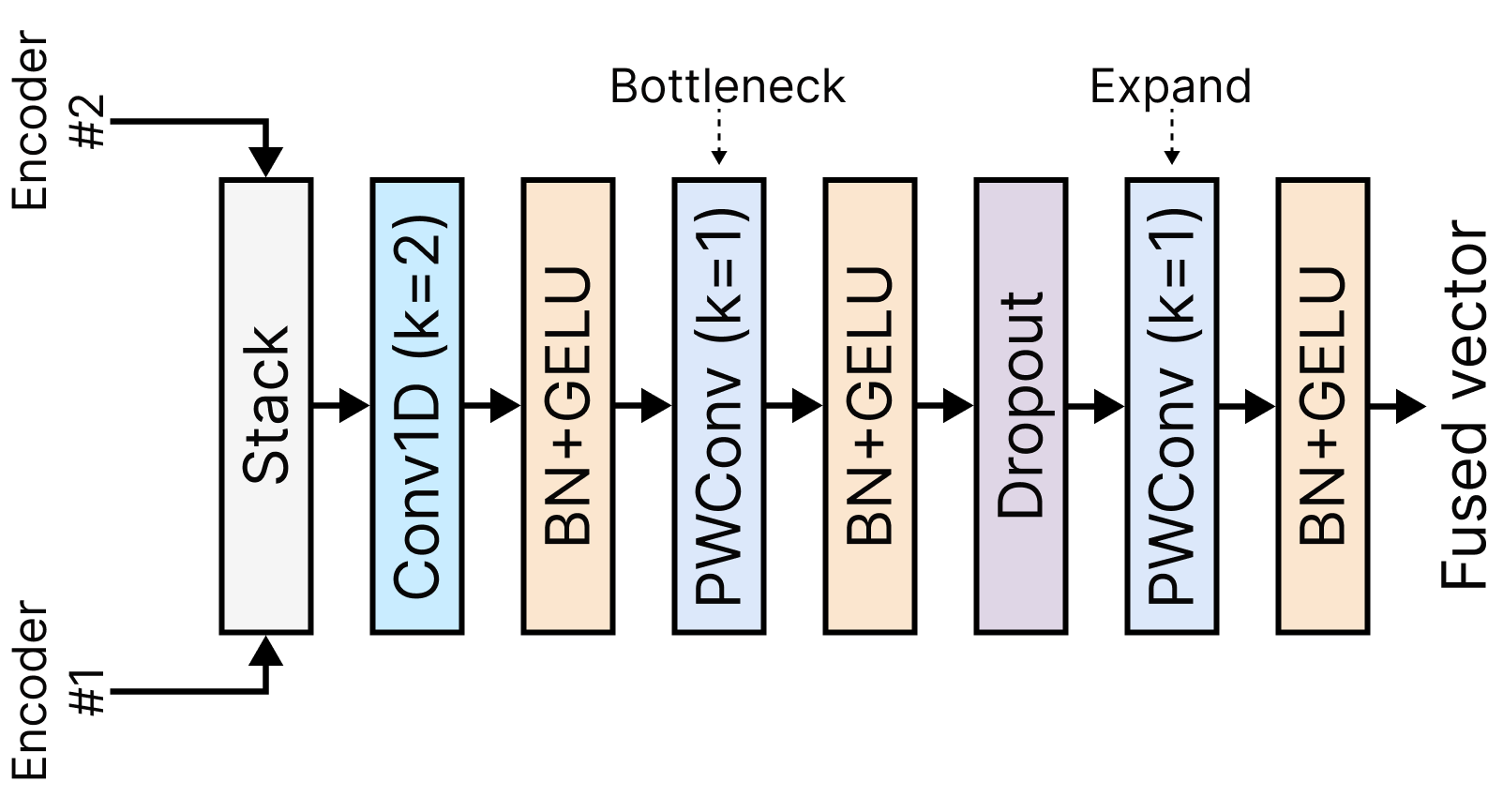}
    \caption{Structure of fusion decoder developed in this study.}
    \label{fig:decoder}
\end{figure}

A custom fusion module (Fig. \ref{fig:decoder}) is designed to combine the outputs of the two encoder branches into a unified representation for classification. The module applies a sequence of 1D convolutional layers along a synthetic sequence dimension of length two, obtained by stacking the encoder outputs. An initial convolution merges the paired feature vectors by learning local interactions between them, followed by a bottleneck formed by two pointwise (kernel size 1) convolutions. The first projects the features to a configurable hidden dimension, while the second restores them to the original size.

More in depth, fusion is performed over a synthetic sequence obtained by stacking the encoder embeddings, $[B,C] \rightarrow [B,C,2]$. A single 1D convolution with kernel size $k=2$ collapses this axis to length one. This operation implements a learned, channel-mixing combination of the two branch embeddings, capturing cross-branch and cross-channel interactions rather than a simple per-channel sum. BatchNorm and GELU stabilize and gate this merge, yielding a fused vector in $[B,C]$.

After fusion, a bottleneck transformation refines the representation using two pointwise convolutions with an optional hidden width $H(C\rightarrow H \rightarrow C)$, each followed by normalization and GELU, with dropout applied between layers for regularization. These pointwise operations act as linear projections on the fused vector, where the bottleneck width $H$ provides a direct mechanism to control model capacity and FLOPs without modifying the encoders. Since the initial $k=2$ merge already collapses the paired features $[f_1, f_2]$ back to width $C$, all subsequent transformations operate at $C$ or at most $H$, preserving low latency. This design offers concrete advantages over common alternatives.

\begin{itemize}
    \item \textbf{Concatenation + MLP:} Concatenating $[f_1, f_2]$ doubles the input width to $2C$; any downstream MLP must first process that larger vector. By merging with a $k=2$ conv, we reduce to $C$ before the non-linear block, so comparable expressive power is achieved with fewer parameters and lower memory traffic.
    \item \textbf{Element-wise sum:} Summation has zero parameters but cannot learn cross-channel reweighting or asymmetric branch contributions. The $k=2$ conv learns both positive/negative mixing and cross-channel couplings, so it can emphasize complementary cues from each branch rather than averaging them away.
    \item \textbf{Attention/gated fusion:} With only two `tokens,' self-attention degenerates to learning a 2$\times$2 mixing plus projections, yet still incurs Q/K/V projections, softmax, and extra activations. The proposed $k=2$ conv realizes an equivalent linear mixing of the two vectors, and the subsequent pointwise + GELU stack supplies non-linearity, achieving similar effect at a fraction of the overhead and with deterministic latency.
\end{itemize}

Two additional properties are purposeful. First, the fusion block is order-aware, as a fixed stacking order is preserved, allowing the model to treat the two encoder branches differently if they specialize during training. Second, the bottleneck/dropout/BN sequence improves optimization stability and calibration without incurring a noticeable speed penalty. %Overall, the fusion module offers a compact and learnable mechanism to model interactions between the encoder embeddings and within the fused representation, providing complementary features to the classifier with minimal parameter and runtime overhead.

\subsection{Training and implementation details}

The models were implemented in PyTorch and trained with mixed-precision using Automatic Mixed Precision (AMP). Optimization was performed with AdamW (learning rate $1 \times 10^{-4}$, weight decay $1 \times 10^{-5}$) and a ReduceLROnPlateau scheduler. All models were trained for 20 epochs with a batch size of 80 using CrossEntropy loss with label smoothing (0.1). To ensure fair comparison, StrokeNeXt and all baselines followed the same training protocol, and all ConvNeXt variants used their default torchvision hyperparameters\footnote{\url{https://docs.pytorch.org/vision/main/models/convnext.html}}. CT images were processed as 3-channel inputs, and all experiments were conducted on a single NVIDIA A100 GPU with 40 GB.

Model evaluation used a comprehensive set of complementary metrics. Classification performance was assessed using Accuracy, Precision, Recall, F1-score, AUROC, AUPRC, and Balanced Accuracy. Reliability and error structure were analyzed through Matthews Correlation Coefficient (MCC), Brier Score, and Expected Calibration Error (ECE). Computational efficiency was evaluated via training time, inference latency, throughput, peak GPU memory usage, and FLOPs. In addition, per-class sensitivity, specificity, and support were reported, and McNemar’s paired tests were applied to assess the statistical significance of performance differences. Implementation of all other metrics employed follow standard formulations. 

% Definitions of Accuracy, Precision, Recall, and F1-score are given in Equations \ref{eq:1}–\ref{eq:4}, while all other metrics follow standard formulations.

% \begin{equation}\label{eq:1} 
% \text{Accuracy} = \frac{TP+TN}{TP + FN + TN + FP}
% \end{equation}

% \begin{equation}\label{eq:2} 
% \text{Precision} = \frac{TP}{TP + FP}
% \end{equation}

% \begin{equation}\label{eq:3} 
% \text{Recall} = \frac{TP}{TP + FN}
% \end{equation}

% \begin{equation}\label{eq:4}
%     \text{F1-score} = 2 \times \frac{\text{Precision} \times \text{Recall}}{\text{Precision} + \text{Recall}}
% \end{equation}

% Implementation of all other metrics employed follow standard formulations.
\section{Results and Discussion}\label{sec:III}

\subsection{Classification of stroke presence}

Table \ref{tab:exp1_tab_1} shows the performance of StrokeNeXt variants on stroke presence classification. All configurations achieve consistently strong results, with accuracy and F1-scores exceeding 97\%. Increasing encoder size leads to systematic improvements across all metrics, indicating that larger ConvNeXt backbones yield more robust feature representations. This trend is further reflected in AUROC and AUPRC values approaching perfect discrimination ($>0.99$) for the base and large variants, demonstrating reliable ranking performance under class imbalance. Balanced accuracy closely matches overall accuracy, and the MCC reaches 0.97 for StrokeNeXt-large, confirming strong predictive agreement in imbalanced settings. Calibration quality remains stable across variants, with low Brier scores and small ECE values, indicating well-aligned probabilistic outputs.

\begin{table*}[!ht]
\centering
\caption{Performance of the different StrokeNeXt variants on stroke presence classification (non-stroke vs. stroke).}\label{tab:exp1_tab_1}
\resizebox{0.9\linewidth}{!}{%
\begin{tabular}{p{2.8cm}p{1.2cm}p{1.2cm}p{0.7cm}p{1.4cm}p{1cm}p{1cm}p{1.2cm}p{1cm}p{1.2cm}p{1cm}p{1cm}p{1.8cm}p{1.4cm}p{1.2cm}p{1cm}p{1.1cm}}
\toprule
\textbf{Method} & \textbf{Accuracy } & \textbf{Precision } & \textbf{Recall } & \textbf{F1-score } & \textbf{AUROC} & \textbf{AUPRC} & \textbf{Balanced Acc.} & \textbf{MCC} & \textbf{Brier score} & \textbf{ECE} & \textbf{Latency {\scriptsize (s)}} & \textbf{Throughput {\scriptsize (img/s)}} &  \textbf{Peak GPU {\scriptsize (GB)}} & \textbf{FLOPs {\scriptsize (G)}} & \textbf{Params {\scriptsize (M)}} & \textbf{Train. time {\scriptsize (h)}} \\\midrule
StrokeNeXt-tiny & 0.978 & 0.978 & 0.978 & 0.978 & 0.986 & 0.987 & 0.971 & 0.950 & 0.021 & 0.053 & 0.002 & 570 & 1.417 & 8.977 & 57.6 & 0.087\\ 
StrokeNeXt-small & 0.980 & 0.981 & 0.981 & 0.980 & 0.976 & 0.981 & 0.972 & 0.957 & 0.019 & 0.047 & 0.003 & 349 & 1.747 & 17.474 & 100.8 & 0.126\\ 
StrokeNeXt-base & 0.982 & 0.983 & 0.982 & 0.982 & 0.990 & 0.990 & 0.977 & 0.960 & 0.017 & 0.051 & 0.004 & 222 & 2.641 & 30.853 & 178.5 & 0.161\\ 
StrokeNeXt-large & 0.987 & 0.987 & 0.987 & 0.987 & 0.995 & 0.995 & 0.983 & 0.970 & 0.013 & 0.059 & 0.009 & 113 & 4.950 & 68.940 & 399.9 & 0.245\\ 
\bottomrule
\end{tabular}}
\end{table*}

Regarding efficiency, StrokeNeXt-tiny exhibits the lowest computational cost, with $<$9 GFLOPs, 58M parameters, and a peak memory usage of 1.4 GB, resulting in a latency of 2 ms per image and a throughput of 570 images per second. As encoder size grows, computational and memory demands increase. StrokeNeXt-large reaches nearly 400M parameters and 69 GFLOPs, with latency rising to 9 ms and throughput dropping to 113 images per second. Despite this, training time remains modest across all variants, with the largest model completing within 15 minutes on the target hardware. This shows that while larger backbones provide incremental accuracy gains, the smaller StrokeNeXt variants already achieve a favorable balance between performance and efficiency.%, particularly under resource or real-time constraints.

Table \ref{tab:exp1_tab_2} compares StrokeNeXt with representative CNN and Transformer baselines, considering both the best-performing and the lightest variant. Conventional CNNs, including MobileNetV2, VGG16, and ResNet variants, reach accuracies in the 86-89\% range with MCC values of 0.65-0.75, indicating frequent misclassifications. Swin Transformer shows improved calibration and discrimination but remains below 90\% accuracy. In contrast, StrokeNeXt-tiny achieves 97.8\% accuracy and an MCC of 0.95, while StrokeNeXt-large reaches 98.7\% accuracy and 0.97 MCC. Both variants attain near-perfect AUROC and AUPRC ($>0.98$), Brier scores $<0.03$, and ECE values $\approx$0.05, demonstrating not only superior classification performance but also better-calibrated probability estimates. %These gains are consistent across metrics, highlighting the advantage of StrokeNeXt design over established CNN and Transformer architectures.

\begin{table*}[!ht]
\centering
\caption{Performance of StrokeNeXt on stroke presence classification (non-stroke vs. stroke) compared with other methods.}\label{tab:exp1_tab_2}
\resizebox{0.9\linewidth}{!}{%
\begin{tabular}{p{3.2cm}p{1.2cm}p{1.2cm}p{0.7cm}p{1.4cm}p{1cm}p{1cm}p{1.2cm}p{1cm}p{1.2cm}p{1cm}p{1cm}p{1.8cm}p{1.4cm}p{1.2cm}p{1cm}p{1.1cm}}
\toprule
\textbf{Method} & \textbf{Accuracy } & \textbf{Precision } & \textbf{Recall } & \textbf{F1-score } & \textbf{AUROC} & \textbf{AUPRC} & \textbf{Balanced Acc.} & \textbf{MCC} & \textbf{Brier score} & \textbf{ECE} & \textbf{Latency {\scriptsize (s)}} & \textbf{Throughput {\scriptsize (img/s)}} &  \textbf{Peak GPU {\scriptsize (GB)}} & \textbf{FLOPs {\scriptsize (G)}} & \textbf{Params {\scriptsize (M)}} & \textbf{Train. time {\scriptsize (h)}} \\\midrule
MobileNetv2 \citep{mobilenetv2} & 0.865 & 0.863 & 0.865 & 0.862 & 0.938 & 0.894 & 0.828 & 0.685 & 0.121 & 0.154 & 0.001 & 3374 & 0.868 & 0.319 & 2.2 & 0.040\\ 
VGG16 \citep{vgg16} & 0.894 & 0.893 & 0.894 & 0.893 & 0.918 & 0.858 & 0.822 & 0.657 & 0.142 & 0.180 & 0.001 & 1750 & 4.020 & 15.519 & 134.2 & 0.041\\ 
ResNet50 \citep{resnet50} & 0.875 & 0.879 & 0.875 & 0.870 & 0.926 & 0.893 & 0.827 & 0.711 & 0.125 & 0.164 & 0.001 & 2057 & 1.034 & 4.130 & 23.5 & 0.040\\ 
ResNet152 \citep{resnet50} & 0.892 & 0.894 & 0.892 & 0.889 & 0.937 & 0.908 & 0.855 & 0.752 & 0.108 & 0.152 & 0.001 & 1213 & 1.298 & 11.601 & 58.2 & 0.046 \\ 
Swin Transformer \citep{swintrans} & 0.893 & 0.893 & 0.892 & 0.890 & 0.939 & 0.899 & 0.859 & 0.751 & 0.094 & 0.099 & 0.002 & 363 & 2.196 & 10.550 & 87.1 & 0.067\\ 
ConvNeXt-base \citep{convnext} & 0.879 & 0.878 & 0.879 & 0.879 & 0.937 & 0.891 & 0.857 & 0.723 & 0.123 & 0.174 & 0.002 & 444 & 1.947 & 15.425 & 87.5 & 0.049\\
\underline{StrokeNeXt-tiny} & \underline{0.978} & \underline{0.978} & \underline{0.978} & \underline{0.978} & \underline{0.986} & \underline{0.987} & \underline{0.971} & \underline{0.950} & \underline{0.021} & \underline{0.053} & \underline{0.002} & \underline{570} & \underline{1.417} & \underline{8.977} & \underline{57.6} & \underline{0.087}\\ 
\textbf{StrokeNeXt-large} & \textbf{0.987} & \textbf{0.987} & \textbf{0.987} & \textbf{0.987} & \textbf{0.995} & \textbf{0.995} & \textbf{0.983} & \textbf{0.970} & \textbf{0.013} & \textbf{0.059} & \textbf{0.009} & \textbf{113} & \textbf{4.950} & \textbf{68.940} & \textbf{399.9} & \textbf{0.245}\\ 
\bottomrule
\end{tabular}}
\vspace{1mm}
\\\scriptsize{\textit{\scriptsize The best results are in bold, the second best are underlined. Note that we have selected the best and lightest StrokeNeXt model for comparison.}}
\end{table*}

In terms of efficiency, lightweight CNNs such as MobileNetV2 and ResNet50 achieve the highest throughput ($>$2000 img/s) with sub-GB memory usage, but their lower accuracy limits competitiveness. VGG16 and ResNet152 retain fast inference but require substantially higher memory ($>$4 GB and $\sim$1.2 GB, respectively), reducing their suitability in constrained environments. StrokeNeXt-tiny, despite its dual-branch design, requires only 57.6M parameters and 8.98 GFLOPs, with 1.4 GB peak memory and 570 img/s throughput. Compared to ConvNeXt-base, VGG16, and Swin, StrokeNeXt-tiny consistently uses fewer parameters, less memory, and achieves higher throughput at comparable latency. Although training time increases slightly, it remains practical ($<0.1$ h). Conversely, StrokeNeXt-large maximizes accuracy at a substantially higher cost (400M parameters, $\sim$5 GB memory, 113 img/s). Overall, the StrokeNeXt family offers a tunable trade-off, with the tiny variant enabling real-time deployment and the large variant favoring peak performance when resources permit.

Table \ref{tab:mcnemar_stroke_normal} reports McNemar test results comparing StrokeNeXt-large (our best model) with the baselines. In all cases, p-values fall well below the $\alpha = 0.05$ threshold, indicating that the observed performance gains are statistically significant rather than random. StrokeNeXt consistently corrects a substantially larger number of cases misclassified by competing models, with at least 67 additional correct predictions across comparisons, while the inverse scenario occurs only rarely. The corresponding $\chi^2$ values further show that these improvements are systematic rather than marginal.

\begin{table}[!ht]
\centering
\caption{Results of the McNemar test comparing our best model with other methods on stroke presence classification (non-stroke vs. stroke).}\label{tab:mcnemar_stroke_normal}
\resizebox{0.85\linewidth}{!}{%
\begin{threeparttable}
\begin{tabular}{p{2.5cm}p{3.2cm}p{1cm}p{1cm}p{0.8cm}p{1.2cm}}
\toprule
\textbf{Method A} & \textbf{Method B} & \textbf{A\cmark B\xmark} & \textbf{B\cmark A\xmark} & \textbf{$\chi^2$} & \textbf{p-value}\\\midrule
StrokeNeXt-large & VGG16 \citep{vgg16} & 94 & 2 & 82.260 & $<$0.0001\\
StrokeNeXt-large & MobileNetv2 \citep{mobilenetv2} & 87 & 4 & 73.890 & $<$0.0001 \\
StrokeNeXt-large & ResNet50 \citep{resnet50} & 79 & 3 & 68.597 & $<$0.0001 \\
StrokeNeXt-large & ResNet152 \citep{resnet50} & 67 & 3 & 56.700 & $<$0.0001 \\
StrokeNeXt-large & ConvNeXt-base \citep{convnext} & 76 & 3 & 65.620 & $<$0.0001 \\
StrokeNeXt-large & Swin Transformer \citep{swintrans} & 69 & 5 & 53.635 & $<$0.0001 \\
\bottomrule
\end{tabular}
  \begin{tablenotes}
    \item A\cmark B\xmark$\!$ indicates the number of samples correctly classified by method A but misclassified by method B, while B\cmark A\xmark$\!$ indicates the opposite. Statistical significance is assessed at $\alpha = 0.05$.
  \end{tablenotes}
  \end{threeparttable}}
\end{table}

Table \ref{tab:per_class_stroke_non_stroke} reports class-wise sensitivity and specificity. Baseline methods exhibit clear imbalances, since ResNet50 and ResNet152 achieve high non-stroke sensitivity ($>0.96$) at the expense of low specificity ($<0.75$), reflecting a tendency to over-predict the non-stroke class. MobileNetV2 and VGG16 show weaker discrimination, with sensitivity and specificity falling below 0.94 in one or both classes. In contrast, StrokeNeXt maintains consistently balanced performance across categories. StrokeNeXt-large achieves the highest overall values, with sensitivity and specificity $>0.99$ for both classes, indicating near-perfect separation, while StrokeNeXt-tiny sustains values around 0.95, outperforming heavier baselines such as ConvNeXt-base and Swin. Unlike models that favor one metric at the expense of the other, StrokeNeXt preserves symmetry, avoiding the trade-offs observed in ResNet and VGG architectures.

\begin{table}[!ht]
\centering
\caption{Per-class performance of StrokeNeXt on stroke presence classification (non-stroke vs. stroke) compared to other methods.}\label{tab:per_class_stroke_non_stroke}
\resizebox{0.75\linewidth}{!}{%
\begin{tabular}{p{3.2cm}p{1.2cm}p{1.4cm}p{1.2cm}p{1.4cm}}
\toprule
\multirow{2}{*}{\textbf{Method}} & \multicolumn{2}{c}{\textbf{Non-stroke}} & \multicolumn{2}{c}{\textbf{Stroke}}\\
\cmidrule(lr){2-3}\cmidrule(lr){4-5}
& Sensitivity & Specificity & Sensitivity & Specificity\\
\midrule
MobileNetv2 \citep{mobilenetv2} & 0.934 & 0.722 & 0.722 & 0.934\\ 
VGG16 \citep{vgg16} & 0.908 & 0.735 & 0.735 & 0.908\\ 
ResNet50 \citep{resnet50} & 0.967 & 0.686 & 0.686 & 0.967\\ 
ResNet152 \citep{resnet50} & 0.965 & 0.744 & 0.744 & 0.965\\ 
Swin Transformer \citep{swintrans} & 0.956 & 0.762 & 0.762 & 0.956\\ 
ConvNeXt-base \citep{convnext} & 0.921 & 0.794 & 0.794 & 0.921\\
StrokeNeXt-tiny & 0.991 & 0.951 & 0.951 & 0.991\\
StrokeNeXt-small & 0.998 & 0.946 & 0.946 & 0.998\\
StrokeNeXt-base & 0.993 & 0.960 & 0.960 & 0.993\\
StrokeNeXt-large & 0.993 & 0.973 & 0.973 & 0.993\\
\bottomrule
\end{tabular}}
\end{table}

% These results show that Strokenext's design improves not only average performance but also balanced recognition across classes, reducing bias toward either category. Consistently higher MCC and lower ECE values further indicate improved robustness under class imbalance and better probability calibration compared to baselines. Overall, the gains of StrokeNeXt extend beyond raw accuracy to more reliable and higher-quality decision making across classes.%potencial a eliminar.

% For a more detailed assessment, Fig. \ref{fig:exp1_fig_1} shows the confusion matrices for StrokeNeXt and the baselines. StrokeNeXt achieves nearly perfect separation, with only a handful of misclassifications in either class, reflecting its high MCC. In contrast, baselines such as ResNet50, VGG16, and MobileNetv2 exhibit substantially higher false negatives in the stroke category, which is particularly problematic in this clinical context. Even more advanced backbones like ConvNeXt-base, Swin Transformer, and ResNet152 show a larger imbalance between false positives and false negatives compared to StrokeNeXt. These results visually confirm the numerical metrics reported earlier, showing that StrokeNeXt not only achieves higher global accuracy and MCC, but also distributes errors more evenly, avoiding the systematic misclassification trends observed in the baselines.

\begin{figure*}[!ht]
    \centering
    \includegraphics[width=0.75\linewidth]{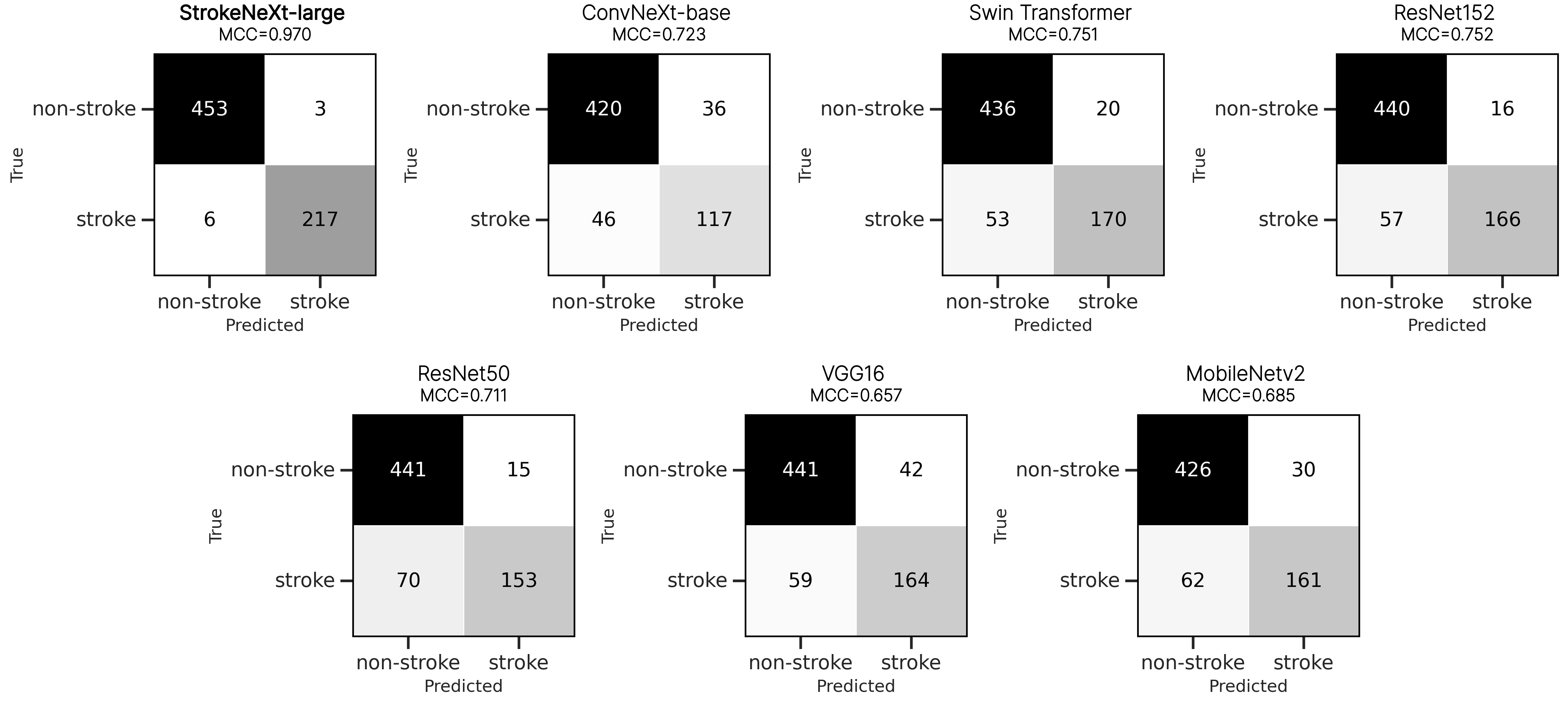}
    \caption{Confusion matrices comparison between StrokeNeXt and other models on stroke presence classification (non-stroke vs. stroke).}
    \label{fig:exp1_fig_1}
\end{figure*}

Fig. \ref{fig:exp1_fig_1} presents the confusion matrices for StrokeNeXt and the baseline models. StrokeNeXt exhibits near-perfect class separation, with only a small number of misclassifications in either category, consistent with its high MCC. In contrast, baselines such as ResNet50, VGG16, and MobileNetV2 show substantially higher false negatives in the stroke class, a particularly critical failure mode in clinical settings. Even stronger backbones, including ConvNeXt-base, Swin Transformer, and ResNet152, display greater imbalance between false positives and false negatives than StrokeNeXt. %Overall, the confusion matrices shows that StrokeNeXt not only improves global accuracy and MCC, but also distributes errors more evenly, avoiding the systematic misclassification patterns observed in others models.

% Next, Table \ref{tab:exp1_tab_3} compares StrokeNeXt with previously reported methods for stroke presence classification. StrokeNeXt-large attains the best overall results, with an F1-score of 0.987, slightly surpassing OzNet (0.984) and clearly outperforming earlier approaches such as CNNs (0.944), StrokeViT (0.870), and classical classifiers like ElasticNet (0.750) and CAD systems (0.730). Importantly, even the lightweight StrokeNeXt-tiny achieves an F1-score of 0.978, outperforming MobileNetv2 (0.974) and approaching the strongest SOTA deep models. This shows that the dual-branch ConvNeXt design scales effectively, as larger variants push performance to the very top of the literature, while smaller variants still provide competitive accuracy at lower computational cost. Overall, the results demonstrate that StrokeNeXt offers a family of models that can flexibly address different trade-offs between efficiency and predictive performance while maintaining superiority over prior work.

Table \ref{tab:exp1_tab_3} compares StrokeNeXt against prior methods for stroke presence classification. StrokeNeXt-large achieves the best overall performance, reaching an F1-score of 0.987, slightly surpassing OzNet (0.984) and clearly outperforming earlier CNN approaches (0.944), StrokeViT (0.870), and classical methods such as ElasticNet (0.750) and CAD systems (0.730). The lightweight StrokeNeXt-tiny also attains an F1-score of 0.978, exceeding MobileNetV2 (0.974) and approaching the strongest models. This indicates that StrokeNeXt scales effectively, with larger variants advancing the state-of-the-art while smaller variants retain competitive accuracy at substantially lower computational cost.

\begin{table}[!ht]
\centering
\caption{Comparison of StrokeNeXt on stroke presence classification (non-stroke vs. stroke) with other methods from the literature.}\label{tab:exp1_tab_3}
\resizebox{0.8\linewidth}{!}{%
\begin{tabular}{p{2.7cm}p{1.4cm}p{1.4cm}p{1.2cm}p{1.4cm}}
\toprule
\textbf{Method} & \textbf{Accuracy} & \textbf{Precision} & \textbf{Recall} & \textbf{F1-score}\\\midrule
%Random Forest \citep{DEV2022100032} & 0.740 & 0.740 & 0.730 & 0.730\\
StrokeViT \citep{RAJ2023105772} & 0.870 & 0.870 & 0.870 & 0.870\\
OzNet \citep{bioengineering9120783} & 0.984 & 0.983 & 0.986 & 0.984\\
CNN \citep{kaya_muh2023} & 0.951 & 0.953 & 0.936 & 0.944\\
% ResNet101 \citep{lo_hung_lin_2021} & 0.809 & - & - & -\\
ElasticNet \citep{DEV2022100032} & 0.760 & 0.790 & 0.710 & 0.750\\
CAD \citep{TURSYNOVA20231431} & 0.810 & 0.760 & 0.820 & 0.730\\ 
MobileNetv2 \citep{YALCIN2022105941} & 0.975 & 0.977 & 0.972 & 0.974\\
\underline{StrokeNeXt-tiny} & \underline{0.978} & \underline{0.978} & \underline{0.978} & \underline{0.978}\\
\textbf{StrokeNeXt-large} & \textbf{0.987} & \textbf{0.987} & \textbf{0.987} & \textbf{0.987}\\ 
\bottomrule
\end{tabular}}
\vspace{1mm}
\\\scriptsize{\textit{\scriptsize The best results are in bold, the second best are underlined.}}
\end{table}

\subsection{Classification of stroke subtypes}

Turning to stroke subtype classification, Table \ref{tab:exp2_tab_1} reports the performance of the StrokeNeXt variants. All models operate close to ceiling, with accuracy and F1-scores in the 0.986-0.988 range and AUROC/AUPRC values between 0.999 and 1.000. Balanced accuracy closely matches overall accuracy, and MCC remains stable at $\approx$0.973, showing consistent behavior across stroke types. Calibration metrics remain low, with small Brier scores and ECE values, indicating well-calibrated probability estimates rather than overconfident predictions. Recall values around 0.985-0.988 imply very few false negatives for either subtype, which is clinically relevant given the divergence in treatment strategies. %The limited variation across variants suggests that dual-branch feature extraction is the primary contributor, with increasing backbone capacity yielding diminishing returns for this task.

\begin{table*}[!ht]
\centering
\caption{Performance of the different StrokeNeXt variants on stroke type classification (ischemia vs. hemorrhage).}\label{tab:exp2_tab_1}
\resizebox{0.9\linewidth}{!}{%
\begin{tabular}{p{2.8cm}p{1.2cm}p{1.2cm}p{0.7cm}p{1.4cm}p{1cm}p{1cm}p{1.2cm}p{1cm}p{1.2cm}p{1cm}p{1cm}p{1.8cm}p{1.4cm}p{1.2cm}p{1cm}p{1.1cm}}
\toprule
\textbf{Method} & \textbf{Accuracy } & \textbf{Precision } & \textbf{Recall } & \textbf{F1-score } & \textbf{AUROC} & \textbf{AUPRC} & \textbf{Balanced Acc.} & \textbf{MCC} & \textbf{Brier score} & \textbf{ECE} & \textbf{Latency {\scriptsize (s)}} & \textbf{Throughput {\scriptsize (img/s)}} &  \textbf{Peak GPU {\scriptsize (GB)}} & \textbf{FLOPs {\scriptsize (G)}} & \textbf{Params {\scriptsize (M)}} & \textbf{Train. time {\scriptsize (h)}} \\\midrule
StrokeNeXt-tiny & 0.986 & 0.985 & 0.986 & 0.986 & 1.000 & 1.000 & 0.986 & 0.973 & 0.011 & 0.058 & 0.002 & 571 & 1.417 & 8.977 & 57.6 & 0.041\\ 
StrokeNeXt-small & 0.987 & 0.987 & 0.987 & 0.987 & 0.999 & 0.999 & 0.986 & 0.973 & 0.013 & 0.049 & 0.003 & 349 & 1.747 & 17.474 & 100.8 & 0.096\\ 
StrokeNeXt-base & 0.988 & 0.988 & 0.988 & 0.988 & 1.000 & 1.000 & 0.986 & 0.973 & 0.011 & 0.054 & 0.005 & 219 & 2.641 & 30.853 & 178.5 &  0.101\\ 
StrokeNeXt-large & 0.987 & 0.987 & 0.985 & 0.986 & 0.999 & 0.999 & 0.986 & 0.973 & 0.015 & 0.061 & 0.009 & 114 & 4.950 & 68.940 & 399.9 & 0.122\\ 
\bottomrule
\end{tabular}}
\end{table*}

Regarding efficiency, StrokeNeXt-tiny remains the most efficient configuration, with 0.002 s latency, 571 img/s throughput, $\approx$58M parameters, and $<$1.5 GB peak GPU memory. In contrast, StrokeNeXt-large incurs higher cost, reaching 0.009 s latency, 114 img/s throughput, $\approx$400M parameters, and a $\sim$5 GB memory footprint. Training time increases with model capacity but remains $<$0.15 h per epoch across variants. Given the marginal performance differences, smaller configurations such as tiny and small offer particularly favorable trade-offs, delivering reliable subtype classification while remaining suitable for deployment.% under inference-time and hardware constraints.

% Table \ref{tab:exp2_tab_2} presents the comparison of StrokeNeXt with other DL models. Here, StrokeNeXt demonstrates a clear advantage over all baselines. Conventional single-branch models such as MobileNetv2, ResNet50, and Swin Transformer show accuracy below 0.86, with corresponding MCC values under 0.75, which signals limited reliability for clinical decision-making. In contrast, both StrokeNeXt-tiny and StrokeNeXt-base reach near-perfect levels across metrics, with AUROC and AUPRC values of 1.0, balanced accuracy of 0.986, and MCC of 0.973, indicating not only excellent overall accuracy but also robustness in handling both classes equally. This improvement is particularly relevant for this application as distinguishing between ischemic and hemorrhagic stroke is critical for treatment, and therapeutic strategies differ substantially depending on the subtype. The low Brier score (0.011) and reduced calibration error (ECE $<$0.06) further suggest that StrokeNeXt’s probability estimates are well-calibrated, increasing confidence in the predictions. Importantly, even the lighter StrokeNeXt-tiny variant achieves the same diagnostic reliability as StrokeNeXt-base, offering flexibility for deployment under different computational constraints without compromising diagnostic quality.

Table \ref{tab:exp2_tab_2} compares StrokeNeXt with other models for stroke subtype classification. Conventional single-branch architectures, (MobileNetV2, ResNet50, and Swin Transformer) achieve accuracy $<0.86$ with MCC values $<0.75$, indicating limited reliability for clinical use. In contrast, StrokeNeXt-tiny and StrokeNeXt-base reach near-ceiling performance, with AUROC/AUPRC of 1.0, balanced accuracy of 0.986, and MCC of 0.973, reflecting robust discrimination across stroke subtypes. Calibration quality is also improved, with a low Brier score (0.011) and ECE $<0.06$. Given the importance of correctly distinguishing stroke subtypes for treatment selection, these gains are clinically meaningful. Notably, StrokeNeXt-tiny matches the diagnostic reliability of StrokeNeXt-base while requiring fewer computational resources.%, enabling flexible deployment without compromising predictive quality.

\begin{table*}[!ht]
\centering
\caption{Performance of StrokeNeXt on stroke type classification (ischemia vs. hemorrhage) compared with other methods.}\label{tab:exp2_tab_2}
\resizebox{0.9\linewidth}{!}{%
\begin{tabular}{p{3.2cm}p{1.2cm}p{1.2cm}p{0.7cm}p{1.4cm}p{1cm}p{1cm}p{1.2cm}p{1cm}p{1.2cm}p{1cm}p{1cm}p{1.8cm}p{1.4cm}p{1.2cm}p{1cm}p{1.1cm}}
\toprule
\textbf{Method} & \textbf{Accuracy } & \textbf{Precision } & \textbf{Recall } & \textbf{F1-score } & \textbf{AUROC} & \textbf{AUPRC} & \textbf{Balanced Acc.} & \textbf{MCC} & \textbf{Brier score} & \textbf{ECE} & \textbf{Latency {\scriptsize (s)}} & \textbf{Throughput {\scriptsize (img/s)}} &  \textbf{Peak GPU {\scriptsize (GB)}} & \textbf{FLOPs {\scriptsize (G)}} & \textbf{Params {\scriptsize (M)}} & \textbf{Train. time {\scriptsize (h)}} \\\midrule
MobileNetv2 \citep{mobilenetv2} & 0.812 & 0.813 & 0.811 & 0.811 & 0.919 & 0.919 & 0.811 & 0.624 & 0.122 & 0.083 & 0.001 & 2341 & 0.868 & 0.319 & 2.2 & 0.020\\ 
VGG16 \citep{vgg16} & 0.879 & 0.879 & 0.879 & 0.879 & 0.951 & 0.956 & 0.879 & 0.758 & 0.104 & 0.124 & 0.001 & 1434 & 4.020 & 15.519 & 134.3 & 0.020\\ 
ResNet50 \citep{resnet50} & 0.852 & 0.854 & 0.852 & 0.852 & 0.918 & 0.924 & 0.851 & 0.706 & 0.133 & 0.136 & 0.001 & 1900 & 1.034 & 4.130 & 23.5 & 0.024\\ 
ResNet152 \citep{resnet50} & 0.865 & 0.873 & 0.865 & 0.865 & 0.949 & 0.936 & 0.864 & 0.738 & 0.101 & 0.130 & 0.001 & 1217 & 1.298 & 11.601 & 58.2 & 0.025\\ 
Swin Transformer \citep{swintrans} & 0.825 & 0.825 & 0.825 & 0.825 & 0.905 & 0.911 & 0.825 & 0.650 & 0.132 & 0.084 & 0.003 & 335 & 2.196 & 10.550 & 87.1 & 0.029\\ 
ConvNeXt-base \citep{convnext} & 0.848 & 0.848 & 0.848 & 0.848 & 0.926 & 0.930 & 0.848 & 0.695 & 0.130 & 0.142 & 0.002 & 441 & 1.947 & 15.425 & 87.6 & 0.026\\ 
\underline{StrokeNeXt-tiny} & \underline{0.986} & \underline{0.985} & \underline{0.986} & \underline{0.986} & \underline{1.000} & \underline{1.000} & \underline{0.986} & \underline{0.973} & \underline{0.011} & \underline{0.058} & \underline{0.002} & \underline{571} & \underline{1.417} & \underline{8.977} & \underline{57.6} & \underline{0.041}\\
\textbf{StrokeNeXt-base} & \textbf{0.988} & \textbf{0.988} & \textbf{0.988} & \textbf{0.988} & \textbf{1.000} & \textbf{1.000} & \textbf{0.986} & \textbf{0.973} & \textbf{0.011} & \textbf{0.054} & \textbf{0.005} & \textbf{219} & \textbf{2.641} & \textbf{30.853} & \textbf{178.5} &  \textbf{0.101}\\ 
\bottomrule
\end{tabular}}
\vspace{1mm}
\\\scriptsize{\textit{\scriptsize The best results are in bold, the second best are underlined. Note that we have selected the best and lightest StrokeNeXt model for comparison.}}
\end{table*}

In terms of efficiency, lightweight baselines such as MobileNetV2 and ResNet50 achieve the lowest FLOPs, parameter counts, and memory usage, yielding high throughput (e.g., $>1900$ img/s for MobileNetV2), but with reduced predictive reliability. At higher capacity, StrokeNeXt-base requires 178.5M parameters and 30.8 GFLOPs, yet maintains competitive latency (0.005 s) and throughput (219 img/s). StrokeNeXt-tiny represents an effective middle ground, with 57.6M parameters and 8.98 GFLOPs, sustaining 571 img/s and 0.002 s latency while outperforming all baselines in accuracy. %This indicates that StrokeNeXt supports efficient deployment in resource-constrained settings, while larger variants naturally scale to scenarios prioritizing maximum reliability over speed.

Table \ref{tab:mcnemar_ischemia_hemorrhage} reports the McNemar test results comparing StrokeNeXt with baseline methods for stroke subtype classification. In all comparisons, p-values fall well below the 0.05 threshold, confirming that the observed gains are statistically significant. StrokeNeXt consistently corrects a larger number of errors made by competing models, while the opposite cases remain minimal. This indicates not only higher overall accuracy, but also more reliable discrimination between ischemia and hemorrhage, a distinction with direct clinical impact due to differing treatment strategies. %Overall, these results support StrokeNeXt as a dependable decision support approach when moving from binary stroke detection to clinically actionable subtype classification.

\begin{table}[!ht]
\centering
\caption{Results of the McNemar test comparing our best model with other methods on stroke type classification (ischemia vs. hemorrhage).}\label{tab:mcnemar_ischemia_hemorrhage}
\resizebox{0.85\linewidth}{!}{%
\begin{threeparttable}
\begin{tabular}{p{2.5cm}p{3.2cm}p{1cm}p{1cm}p{0.8cm}p{1.2cm}}
\toprule
\textbf{Method A} & \textbf{Method B} & \textbf{A\cmark B\xmark} & \textbf{B\cmark A\xmark} & \textbf{$\chi^2$} & \textbf{p-value}\\\midrule
StrokeNeXt-base & VGG16 \citep{vgg16} & 25 & 1 & 20.346 & $<$0.0001\\
StrokeNeXt-base & MobileNetv2 \citep{mobilenetv2} & 40 & 1 & 35.219 & $<$0.0001 \\
StrokeNeXt-base & ResNet50 \citep{resnet50} & 31 & 1 & 26.281 & $<$0.0001 \\
StrokeNeXt-base & ResNet152 \citep{resnet50} & 29 & 2 & 21.807 & $<$0.0003 \\
StrokeNeXt-base & ConvNeXt-base \citep{convnext} & 32 & 1 & 27.272 & $<$0.0001 \\
StrokeNeXt-base & Swin Transformer \citep{swintrans} & 37 & 1 & 32.236 & $<$0.0001 \\
\bottomrule
\end{tabular}
  \begin{tablenotes}
    \item A\cmark B\xmark$\!$ indicates the number of samples correctly classified by method A but misclassified by method B, while B\cmark A\xmark$\!$ indicates the opposite. Statistical significance is assessed at $\alpha = 0.05$.
  \end{tablenotes}
  \end{threeparttable}}
\end{table}

Table \ref{tab:per_class_ischemia} reports per-class performance for ischemic and hemorrhagic stroke classification. Conventional models show moderate results, with ischemia sensitivity often below 0.90, a concerning limitation given the clinical risk of missed ischemic cases. In contrast, StrokeNeXt variants achieve higher sensitivity ($>0.97$) while maintaining strong specificity, indicating balanced discrimination across both subtypes. The high ischemia sensitivity reflects improved reliability in identifying the more frequent and diagnostically challenging class, while the comparable performance for hemorrhage confirms cross-category robustness. %Overall, these results support StrokeNeXt as a decision support approach that enables more reliable subtype differentiation than existing deep learning baselines.

\begin{table}[!ht]
\centering
\caption{Per-class performance of StrokeNeXt on troke type classification (ischemia vs. hemorrhage) compared to other methods.}\label{tab:per_class_ischemia}
\resizebox{0.75\linewidth}{!}{%
\begin{tabular}{p{3.2cm}p{1.2cm}p{1.4cm}p{1.2cm}p{1.4cm}}
\toprule
\multirow{2}{*}{\textbf{Method}} & \multicolumn{2}{c}{\textbf{Ischemia}} & \multicolumn{2}{c}{\textbf{Hemorrhage}}\\
\cmidrule(lr){2-3}\cmidrule(lr){4-5}
& Sensitivity & Specificity & Sensitivity & Specificity\\
\midrule
MobileNetv2 \citep{mobilenetv2} & 0.773 & 0.850 & 0.850 & 0.772\\ 
VGG16 \citep{vgg16} & 0.873 & 0.885 & 0.885 & 0.873\\ 
ResNet50 \citep{resnet50} & 0.809 & 0.894 & 0.894 & 0.809\\ 
ResNet152 \citep{resnet50} & 0.790 & 0.938 & 0.938 & 0.790\\ 
Swin Transformer \citep{swintrans} & 0.818 & 0.832 & 0.832 & 0.818\\ 
ConvNeXt-base \citep{convnext} & 0.846 & 0.850 & 0.850 & 0.846\\
StrokeNeXt-tiny & 0.972 & 0.999 & 0.999 & 0.972\\
StrokeNeXt-small & 0.982 & 0.991 & 0.991 & 0.982\\
StrokeNeXt-base & 0.973 & 0.999 & 0.999 & 0.973\\
StrokeNeXt-large & 0.982 & 0.991 & 0.991 & 0.982\\
\bottomrule
\end{tabular}}
\end{table}

\begin{figure*}[!ht]
    \centering
    \includegraphics[width=0.75\linewidth]{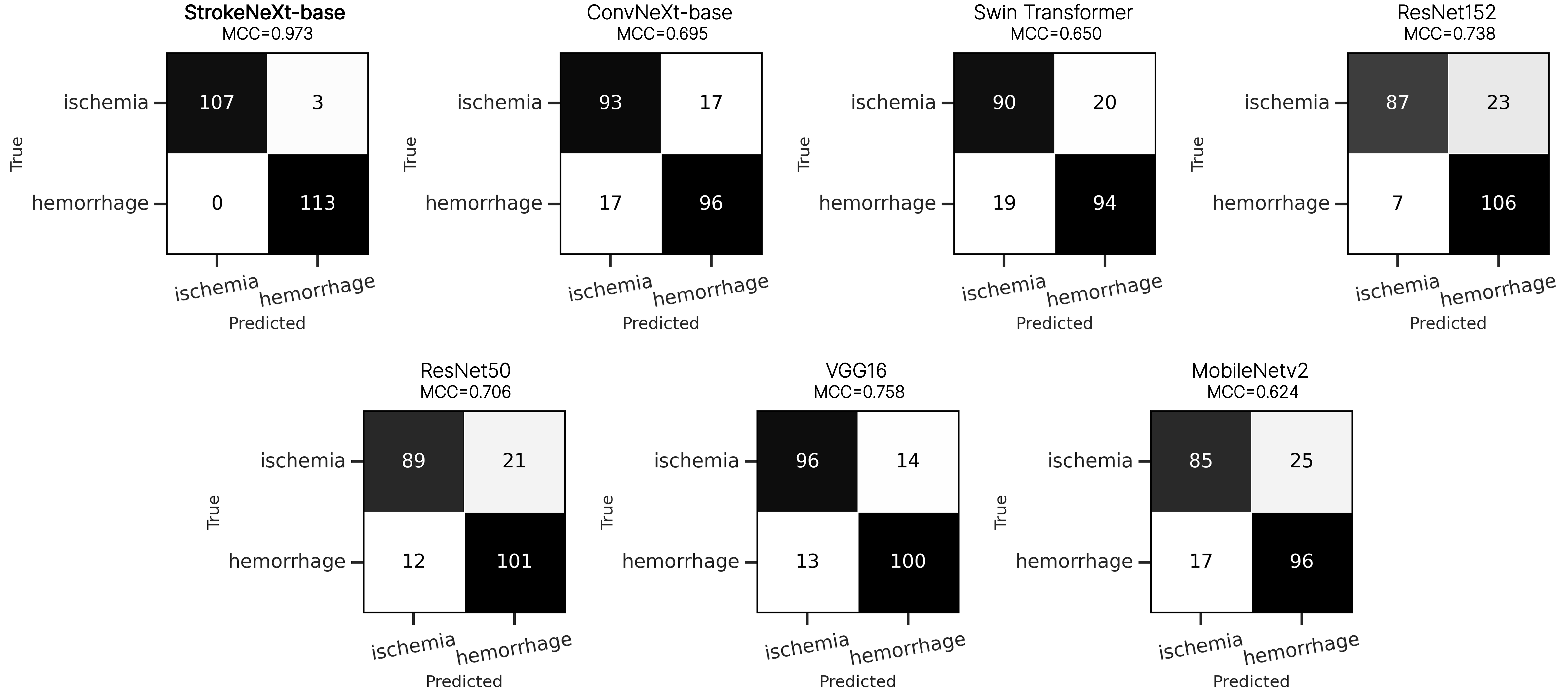}
    \caption{Confusion matrices comparison between StrokeNeXt and other models on stroke type classification (ischemia vs. hemorrhage).}
    \label{fig:exp2_fig_1}
\end{figure*}

Fig. \ref{fig:exp2_fig_1} presents the confusion matrices for stroke subtype classification. StrokeNeXt-base shows near-perfect separation between subtypes, with only a few misclassifications, in contrast to the higher error rates observed in ResNet, Swin Transformer, and MobileNetV2. Notably, StrokeNeXt is the only model correctly identifying more than 100 ischemic cases, while all others fall below this threshold. Concurrently, it achieves perfect hemorrhage recognition, avoiding false negatives that could critically affect clinical decisions. This shows that StrokeNeXt not only improves overall accuracy but also minimizes subtype confusion, a key requirement in settings where treatment depends on precise stroke categorization. Higher MCC values further support its robustness under balanced evaluation.%, while the low calibration error (ECE) indicates reliable probabilistic predictions suitable for clinical decision support.

Table \ref{tab:exp2_tab_3} compares StrokeNeXt against prior methods for ischemic vs. hemorrhagic classification. StrokeNeXt-base achieves the highest numbers, with an F1-score of 0.988, outperforming CNN approaches such as D-UNet (0.985) and P-CNN (0.983). Notably, StrokeNeXt-tiny matches or exceeds these baselines with an F1-score of 0.986 at a lower computational cost. In contrast, Random Forest and Enhanced-CNN remain below 0.96, and the 3D-CNN reaches only 0.880 despite using volumetric information. %Overall, these results show that the dual-branch design surpasses prior strategies while offering scalable configurations, from efficient compact models to variants that push state-of-the-art accuracy.

\begin{table}[!ht]
\centering
\caption{Comparison of StrokeNeXt on stroke type classification (ischemia vs. hemorrhage) with other methods from the literature.}\label{tab:exp2_tab_3}
\resizebox{0.8\linewidth}{!}{%
\begin{tabular}{p{2.9cm}p{1.4cm}p{1.4cm}p{1.2cm}p{1.4cm}}
\toprule
\textbf{Method} & \textbf{Accuracy} & \textbf{Precision} & \textbf{Recall} & \textbf{F1-score}\\\midrule
P-CNN-W \citep{GAUTAM2021102178} & 0.975 & 0.975 & 0.975 & 0.975\\
Random Forest \citep{9179307} & 0.959 & 0.944 & 0.961 & 0.954\\
% ANN \citep{Shakunthala2023} & 0.889 & 0.892 & 0.928 & 0.910\\
Enhanced-CNN \citep{Shakunthala2023} & 0.952 & 0.949 & 0.972 & 0.960\\
P-CNN \citep{GAUTAM2021102178} & 0.983 & 0.983 & 0.983 & 0.983\\
3D-CNN \citep{NEETHI2022103720} & 0.920 & 0.940 & 0.840 & 0.880\\
D-UNet \citep{YALCIN2022105941} & 0.985 & 0.986 & 0.985 & 0.985\\
\underline{StrokeNeXt-tiny} & \underline{0.986} & \underline{0.985} & \underline{0.986} & \underline{0.986}\\
\textbf{StrokeNeXt-base} & \textbf{0.988} & \textbf{0.988} & \textbf{0.988} & \textbf{0.988}\\
\bottomrule
\end{tabular}}
\vspace{1mm}
\\\scriptsize{\textit{\scriptsize The best results are in bold, the second best are underlined.}}
\end{table}

Overall, the results show that the proposed StrokeNeXt achieves higher accuracy and statistical reliability than competing approaches, with improvements consistently supported by significance testing. It maintains strong calibration alongside high predictive performance, ensuring outputs that are both discriminative and trustworthy for clinical decision support. These properties hold across both stroke presence detection and subtype classification, indicating that the architecture generalizes effectively to different diagnostic tasks. Importantly, the efficiency analysis demonstrates that these gains do not compromise usability, as even lightweight StrokeNeXt variants outperform conventional baselines while offering favorable trade-offs between accuracy and computational cost. This flexibility enables deployment across a wide range of resource settings, positioning StrokeNeXt as a reliable and efficient alternative to existing methods for CT-based stroke classification. %Ablation test can be found in the supplemental materials.

\section{Limitations}\label{sec:IV}

This study has limitations that should be acknowledged. First, the dataset does not include patient-level information such as demographics or clinical outcomes. As StrokeNeXt is purely image-based, the absence of these variables limits the analysis of how model predictions may generalize across different patient populations or interact with clinical variables. In addition, the evaluation is restricted to CT imagery, and the reported efficiency results are specific to the hardware configuration used in this work. While these results provide a realistic estimate, validation on larger and more diverse cohorts and on different hardware would be necessary to fully assess robustness and generalizability. in real-world clinical settings.

% This study has certain limitations that should be acknowledged. First, the dataset used does not include patient-level information such as demographics, comorbidities, or clinical outcomes. While our focus was on image-based classification, the absence of these factors limits the possibility of analysing how the model’s predictions might generalize across different populations or interact with clinical variables. Incorporating such data in future studies could provide a more comprehensive assessment of the model’s utility in real-world diagnostic settings.

% Second, our evaluation was restricted to CT scans, and further experiments with other imaging modalities such as MRI could provide additional insights. Exploring multi-modal integration would also help determine whether the proposed dual-branch architecture can leverage complementary sources of information to further enhance diagnostic performance. In addition, although the experimental results were validated with strong statistical evidence, broader validation on larger and more diverse cohorts would strengthen confidence in the model’s robustness across clinical environments.

% Finally, another limitation is that the reported efficiency metrics, including training and inference times as well as memory usage, are tied to the specific hardware employed in this study. While they provide a realistic estimate for the tested configuration, further evaluation on diverse hardware environments would be needed to fully establish generalizability in time-critical clinical settings.

\section{Conclusions}\label{sec:V}

We present StrokeNeXt for brain stroke classification from CT imagery. It is a dual-branch architecture with two identical ConvNeXt encoders operating in parallel on the same input. The extracted features are fused through a lightweight convolutional decoder and processed by a compact classifier, aiming to enhance representational capacity while preserving computational efficiency. The approach is evaluated on a curated dataset of 6,774 annotated CT images under two classification scenarios: stroke presence detection and stroke subtype classification (ischemic vs. hemorrhagic).

Across both tasks, StrokeNeXt consistently outperforms convolutional, Transformer-based, and state-of-the-art methods. StrokeNeXt achieves F1-scores of up to 0.988, while maintaining consistently high class-wise sensitivity and specificity, low calibration error, and strong Matthews correlation coefficients. Statistical significance testing further indicates that the observed improvements are not due to random variation. Confusion matrix analysis also demonstrates reliable discrimination of challenging cases. The model maintains fast inference and practical training times, demonstrating that the approach design effectively balances diagnostic performance and practical deployability.

{
    \clearpage
    \small
    \bibliographystyle{ieeenat_fullname}
    \bibliography{main}

@String(CVPR= {IEEE Conf. Comput. Vis. Pattern Recog.})

@String(ICCV= {Int. Conf. Comput. Vis.})

@String(CVPR  = {CVPR})

@String(ICCV  = {ICCV})

@article{HOSSAIN2025109711,
title = {A novel hybrid ViT-LSTM model with explainable AI for brain stroke detection and classification in CT images: A case study of Rajshahi region},
journal = {Comput. Biol. Med.},
volume = {186},
pages = {109711},
year = {2025},
author = {Md. Maruf Hossain and Md. Mahfuz Ahmed and Abdullah Al Nomaan Nafi and Md. Rakibul Islam and Md. Shahin Ali and Jahurul Haque and Md Sipon Miah and Md Mahbubur Rahman and Md Khairul Islam}
}

@article{ZHU2022147,
title = {Application of Deep Learning to Ischemic and Hemorrhagic Stroke Computed Tomography and Magnetic Resonance Imaging},
journal = {Semin. Ultrasound CT MRI},
volume = {43},
number = {2},
pages = {147-152},
year = {2022},
author = {Guangming Zhu and Hui Chen and Bin Jiang and Fei Chen and Yuan Xie and Max Wintermark}}

@article{GAUTAM2021102178,
title = {Towards effective classification of brain hemorrhagic and ischemic stroke using CNN},
journal = {Biomed. Signal Process. Control.},
volume = {63},
pages = {102178},
year = {2021},
author = {Anjali Gautam and Balasubramanian Raman}}

@inproceedings{10112284,
  author={Saini, Archana and Guleria, Kalpna and Sharma, Shagun},
  booktitle={INDIACom}, 
  title={Performance Analysis of Machine Learning Approaches for Stroke Prediction in Healthcare}, 
  year={2023}}

@article{RAJ2023105772,
title = {StrokeViT with AutoML for brain stroke classification},
journal = {Eng. Appl. Artif. Intell.},
volume = {119},
pages = {105772},
year = {2023},
author = {Rishi Raj and Jimson Mathew and Santhosh Kumar Kannath and Jeny Rajan}}

@article{Shakunthala2023,
author = {M. Shakunthala and K. HelenPrabha},
title ={Classification of ischemic and hemorrhagic stroke using Enhanced-CNN deep learning technique},
journal = {J. Intell. Fuzzy Syst.},
volume = {45},
number = {4},
pages = {6323-6338},
year = {2023}}

@article{kaya_muh2023, title={A CNN transfer learning-based approach for segmentationand classification of brain stroke from noncontrast CTimages}, volume={33}, number={4}, journal={Int. J. Imaging Syst. Technol.}, author={Kaya, Buket and Muhammed Onal}, year={2023}, pages={1335--1352}}

@article{cabral_powers_2022, title={Comparison of Outcomes of Ischemic Stroke Initially Imaged With Cranial Computed Tomography Alone vs Computed Tomography Plus Magnetic Resonance Imaging}, volume={5}, number={7}, journal={JAMA Network Open}, author={Cabral Frade, Heitor and Wilson, Susan E. and Beckwith, Anne and Powers, William J.}, year={2022}, pages={e2219416} }

@article{ABBASI2023100145,
title = {Automatic brain ischemic stroke segmentation with deep learning: A review},
journal = {Neurosci. Inform.},
volume = {3},
number = {4},
pages = {100145},
year = {2023},
author = {Hossein Abbasi and Maysam Orouskhani and Samaneh Asgari and Sara Shomal Zadeh}
}

@article{AGGARWAL2022105350,
title = {COVID-19 image classification using deep learning: Advances, challenges and opportunities},
journal = {Comput. Biol. Med.},
volume = {144},
pages = {105350},
year = {2022},
author = {Priya Aggarwal and Narendra Kumar Mishra and Binish Fatimah and Pushpendra Singh and Anubha Gupta and Shiv Dutt Joshi}}

@article{DESHPANDE2021102573,
title = {Automatic segmentation, feature extraction and comparison of healthy and stroke cerebral vasculature},
journal = {NeuroImage: Clin.},
volume = {30},
pages = {102573},
year = {2021},
author = {Aditi Deshpande and Nima Jamilpour and Bin Jiang and Patrik Michel and Ashraf Eskandari and Chelsea Kidwell and Max Wintermark and Kaveh Laksari}}

@article{CAI2022102522,
title = {DeepStroke: An efficient stroke screening framework for emergency rooms with multimodal adversarial deep learning},
journal = {Med. Image Anal.},
volume = {80},
pages = {102522},
year = {2022},
author = {Tongan Cai and Haomiao Ni and Mingli Yu and Xiaolei Huang and Kelvin Wong and John Volpi and James Z. Wang and Stephen T.C. Wong}}

@inproceedings{10971457,
  author={Ramos, Leo Thomas and Sappa, Angel D.},
  booktitle={SoutheastCon}, 
  title={Dual-Branch ConvNeXt-Based Network with Attentional Fusion Decoding for Land Cover Classification Using Multispectral Imagery}, 
  year={2025}}

@article{10623211,
  author={Ramos, Leo Thomas and Sappa, Angel D.},
  journal={IEEE JSTARS}, 
  title={Multispectral Semantic Segmentation for Land Cover Classification: An Overview}, 
  year={2024},
  volume={17},
  pages={14295-14336}}

@article{esteva_chou21, title={Deep learning-enabled medical computer vision}, volume={4}, number={1}, journal={npj Digit. Med.}, author={Esteva, Andre and Chou, Katherine and Yeung, Serena and Naik, Nikhil and Madani, Ali and Mottaghi, Ali and Liu, Yun and Topol, Eric and Dean, Jeff and Socher, Richard}, year={2021}}

@article{ramos2026multiencoder,
  title={Multi-encoder ConvNeXt Network with Smooth Attentional Feature Fusion for Multispectral Semantic Segmentation},
  author={Leo Thomas Ramos and Angel D. Sappa},
  journal={arXiv preprint arXiv:2602.10137},
  year={2026}
}

@article{ELHARROUSS2024100645,
title = {Backbones-review: Feature extractor networks for deep learning and deep reinforcement learning approaches in computer vision},
journal = {Computer Science Review},
volume = {53},
pages = {100645},
year = {2024},
author = {Omar Elharrouss and Younes Akbari and Noor Almadeed and Somaya Al-Maadeed}}

@inproceedings{9179307,
  author={Badriyah, Tessy and Sakinah, Nur and Syarif, Iwan and Syarif, Daisy Rahmania},
  booktitle={ICECCE}, 
  title={Machine Learning Algorithm for Stroke Disease Classification}, 
  year={2020}}

@article{bioengineering9120783,
AUTHOR = {Ozaltin, Oznur and Coskun, Orhan and Yeniay, Ozgur and Subasi, Abdulhamit},
TITLE = {A Deep Learning Approach for Detecting Stroke from Brain CT Images Using OzNet},
JOURNAL = {Bioengineering},
VOLUME = {9},
pages={1--16},
YEAR = {2022},
NUMBER = {12}}

@article{YALCIN2022105941,
title = {Brain stroke classification and segmentation using encoder-decoder based deep convolutional neural networks},
journal = {Comput. Biol. Med.},
volume = {149},
pages = {105941},
year = {2022},
author = {Sercan Yalcın and Huseyin Vural}}

@article{NEETHI2022103720,
title = {Stroke classification from computed tomography scans using 3D convolutional neural network},
journal = {Biomed. Signal Process. Control.},
volume = {76},
pages = {103720},
year = {2022},
author = {A.S. Neethi and S. Niyas and Santhosh Kumar Kannath and Jimson Mathew and Ajimi Mol Anzar and Jeny Rajan}}

@article{abdou_2022, title={Literature review: efficient deep neural networks techniques for medical image analysis}, volume={34}, journal={Neural Comput. Appl.}, author={Abdou, Mohamed A.}, year={2022}, pages={5791–5812} }

@article{koc_akc2022, title={Artificial Intelligence in Healthcare Competition (Teknofest-2021): Stroke Data Set}, volume={54}, journal={Eurasian J. Med.}, author={Koc, Ural and Akcapinar Sezer, Ebru and Ozkaya, Yasar Alper and Yarbay, Yasin and Taydas, Onur and Ayyildiz, Veysel Atilla and Kiziloglu, Huseyin Alper and Kesimal, Ugur and Cankaya, Imran and Besler, Muhammed Said and Karakas, Emrah and Karademir, Fatih and Sebik, Nihat Baris and Bahadir, Murat and Sezer, Ozgur and Yesilyurt, Batuhan and Varli, Songul and Akdogan, Erhan and Ulgu, Mustafa Mahir and Birinci, Suayip}, year={2022}, pages={248–258} }

@article{TURSYNOVA20231431,
title = {Deep Learning-Enabled Brain Stroke Classification on Computed Tomography Images},
journal = {Comput. Mater. Contin.},
volume = {75},
number = {1},
pages = {1431-1446},
year = {2023},
author = {Azhar Tursynova and Batyrkhan Omarov and Natalya Tukenova and Indira Salgozha and Onergul Khaaval and Rinat Ramazanov and Bagdat Ospanov},
}

@article{DEV2022100032,
title = {A predictive analytics approach for stroke prediction using machine learning and neural networks},
journal = {Healthc. Anal.},
volume = {2},
pages = {100032},
year = {2022},
author = {Soumyabrata Dev and Hewei Wang and Chidozie Shamrock Nwosu and Nishtha Jain and Bharadwaj Veeravalli and Deepu John},
}

@inproceedings{mobilenetv2,
  author={Mark Sandler and Andrew Howard and Menglong Zhu and Andrey Zhmoginov and Liang-Chieh Chen},
  booktitle={CVPR}, 
  title={MobileNetV2: Inverted Residuals and Linear Bottlenecks}, 
  year={2019}}

@inproceedings{resnet50,
  author={Kaiming He and Xiangyu Zhang and Shaoqing Ren and Jian Sun},
  booktitle={CVPR}, 
  title={Deep Residual Learning for Image Recognition}, 
  year={2015}}

@inproceedings{swintrans,
  author={Ze Liu and Yutong Lin and Yue Cao and Han Hu and Yixuan Wei and Zheng Zhang and Stephen Lin and Baining Guo},
  booktitle={ICCV}, 
  title={Swin Transformer: Hierarchical Vision Transformer using Shifted Windows}, 
  year={2021}}

@inproceedings{vgg16,
  author={Liu, Shuying and Deng, Weihong},
  booktitle={ACPR}, 
  title={Very deep convolutional neural network based image classification using small training sample size}, 
  year={2015}}

@inproceedings{convnext,
  author={Zhuang Liu and Hanzi Mao and Chao-Yuan Wu and Christoph Feichtenhofer and Trevor Darrell and Saining Xie},
  booktitle={CVPR}, 
  title={A ConvNet for the 2020s}, 
  year={2022}}
}


\end{document}